\documentclass[aps,twocolumn]{revtex4}
\usepackage{psfig}
\usepackage{subfigure}
\usepackage{graphicx}
\usepackage{amssymb}

\begin{document}



\title{Adaptive synchronization of dynamics on evolving complex networks}
\author{Francesco Sorrentino${}^{\ddagger*}$, Edward Ott${}^{*}$ }
\affiliation{${}^\ddagger$ University of Naples Federico II, Naples 80125, Italy \\ ${}^*$ Institute for Research in Electronics and Applied Physics, Department of Physics, and Department of Electrical and Computer Engineering, University of Maryland, College Park, Maryland 20742}
\begin{abstract}
We study the problem of synchronizing a general complex network
 by means of an adaptive strategy in the case where the network topology is slowly time varying and every node receives at each time only one aggregate signal from the set of its neighbors. We introduce an appropriately defined potential that each node seeks to minimize in order to reach/maintain synchronization. We show that our strategy is effective in tracking synchronization as well as in achieving synchronization when appropriate conditions are met.
\end{abstract}

\maketitle

In recent years synchronization of large networks of interconnected systems has been the subject of intense investigation. In \cite{Pe:Ca} it was shown that, under the assumption that all the systems are identical and the coupling among the connected systems is of a suitable type, the stability of the synchronous evolution can be investigated by means of a `master stability function' approach. 
Many papers have followed the approach in \cite{Pe:Ca}, focusing on the way the network topology impacts the stability of the synchronous evolution \cite{Ba:Pe02}, e.g., reviewed in \cite{Report}. 
An adaptive synchronization approach to obtaining estimates of unknown system parameters has been pursued in a number of recent papers \cite{ADAP1}. 
In \cite{Zh:Ku06} it was shown that an adaptive strategy acting on the strengths of the 
network couplings, based on information about the dynamics at its nodes, can be effective in enhancing the stability of the synchronous state. Here we will present an adaptive approach to synchronize a time varying network 
that evolves under the effects of exogenous (unpredictable) factors. 


The formulation of many past works on network synchronization of identical systems (see \cite{Pe:Ca} and the related literature) typically involves coupling to a node $i$ from other network nodes $j$ through a term of the form
\begin{equation}
\{ \sum_j A_{ij} H (x_j(t)) \} - \{ (\sum_j A_{ij}) H (x_i(t)) \}, \label{previous}
\end{equation}
where $A_{ij}$ is the weighed $N \times N$ network adjacency matrix representing the strength of the coupling from $j$ to $i$, $(A_{ii}=0)$, $x_i(t)$ is the $n$-dimensional state at node $i$,  $H: \mathbf{R}^{n} \rightarrow  \mathbf{R}^{n}$, 
and the network has $N$ nodes, $\{i,j\}=1,2,...,N$. For the purposes of our approach, we think of the first term in (\ref{previous}) as a directly accessible physical signal received by node $i$ from other nodes in the network, and we denote this signal
\begin{equation}
s_i(t)= \sum_j A_{ij} H(x_j(t)). \label{s_i}
\end{equation}
The second term in (\ref{previous}) results in the convenient property that the coupling becomes zero when synchronization is achieved; i.e., when
\begin{equation}
x_1(t)=x_2(t)=...=x_N(t)=x_s(t). \label{si}
\end{equation}
In order to implement a coupling of the form of (\ref{previous}), the external information required at node $i$ is the received signal $s_i(t)$, as well as the sum of the input coupling strengths, $\sum_j A_{ij}$. Here we will be concerned with situations in which the only available external information at node $i$ is the signal $s_i(t)$, and direct knowledge of $\sum_j A_{ij}$ is unavailable. Thus synchronization must be achieved on the basis of $s_i(t)$ only. In order to accomplish this, we will propose and test a simple adaptive strategy. We will also present numerical experiments that show that our adaptive strategy can, under appropriate conditions, be effective in synchronizing the network.

Consideration of this problem has both technological and biological motivation. For example, as a technological motivation, we consider a networked system in which dynamical units (e.g., chaotic oscillators) are located on autonomous moving platforms (nodes) and communicate by wireless. The signal received by each platform is the weighed sum of the signals (represented by $H(x_j)$) sent from other platforms, where the weights represent the spreading and attenuation of these signals along their propagation paths (represented by $A_{ij}$). If at each platform $i$ there is no available information on the individual locations, attenuations, etc., associated with the input from other platforms, then the adaptation of a node to changes in the network due to motion of the platforms must be accomplished solely on the basis of the aggregate signal (\ref{s_i}) that it receives. In the biological context, we note that synchronism is often observed in circumstances of changing environments and states of the considered organism (e.g., the ability to synchronize is evidently evolved as an organism develops, and persists with changes due to disease, etc.). In both the above technological example, as well as possible biological examples, in order to maintain overall system synchronism, nodes must adjust the processing of inputs they receive from the network based only on aggregate information available to them. Furthermore, in both the technological and biological contexts, we envision that 
synchronous dynamics on which the node states $x_i(t)$ evolve is typically much faster than the time scale over which the network changes. Thus, we say that the network changes `slowly' in time, and we will make use of this supposed slowness in what follows. To emphasize this, we will sometimes write the adjacency matrix as $A_{ij}(t)$ instead of $A_{ij}$.
We consider network dynamical equations of the following form,
\begin{equation}
\dot x_i(t) =  F(x_i(t)) + \sigma_i(t) s_i(t) - \gamma H (x_i(t)),  \quad i=1,2,...,N, \label{main}
\end{equation}
where the equation for the evolution of $x_i(t)$ in the absence of coupling is $\dot x_i(t)=F(x_i(t))$ with
$F: \mathbf{R}^{n} \rightarrow  \mathbf{R}^{n}$, $s_i(t)$ is the input signal at node $i$ defined in Eq. (\ref{s_i}), and $\gamma$ is a constant gain equal for all the nodes in the network.

A synchronous dynamical solution (\ref{si}) exists for $\sigma_i(t)$ equal to
\begin{equation}
\bar \sigma_i(t)= \gamma/{\sum_j A_{ij}(t)}. \label{cond}
\end{equation}
When the condition
\begin{equation}
\sigma_i(t)=\bar{\sigma}_i(t) \label{sigma}
\end{equation}
is satisfied, Eq. (\ref{main}) can be rewritten as
\begin{equation}
\dot x_i(t) =  F(x_i(t)) + \gamma  \sum_j \mathcal{L}_{ij}(t) H({x}_j(t)),  \label{Lapl}
\end{equation}
where the $N \times N$ matrix $\mathcal{L}(t)=\{\mathcal{L}_{ij}(t) \}$ is such that $\mathcal{L}_{ii}= -1$ $\forall i$, $\mathcal{L}_{ij}(t)= A_{ij}(t)/\sum_j A_{ij}(t)$, for $i \neq j$ and thus has the property that the sum of the elements in each row is zero. Thus, assuming synchronization, $x_1(t)=x_2(t)=...=x_N(t)=x_s(t)$, the last term in (\ref{Lapl}) is identically zero and the synchronization dynamics is governed by the same dynamics as for an individual uncoupled system,
\begin{equation}
\dot x_s(t) =  F(x_s(t)).  \label{xs}
\end{equation}

In this situation, the arguments of the master stability function theory of Ref. \cite{Pe:Ca} apply and the stability of the synchronous evolution depends essentially on the choice of an appropriate coupling $\gamma$. When $\sigma_i(t)$ is not given by (\ref{cond}), then Eq. (\ref{Lapl}) does not in general admit a synchronous solution.

In what follows we will attempt to program the time evolution of $\sigma_i(t)$ so that it tends to relax toward $\bar \sigma_i(t)$, and we will proceed under the assumption that, for the chosen value of $\gamma$, the synchronized state is stable for $\sigma_i= \bar \sigma_i$. In order to motivate our programming technique, we first define a mean squared exponentially weighed synchronization error at each node $i$,
\begin{equation}
\tilde{\triangle}_i(t)=  \int^t e^{-\nu (t-t')} {| \sigma_i(t') s_i(t')- \gamma H (x_i(t')) |}^2 dt', \label{tri}
\end{equation}
where $\nu^{-1}$ is the temporal extent over which the averaging is performed. 
 Thus when synchronization is achieved (i.e., $x_i=x_s$ and $\sigma_i=\bar \sigma_i$ for all $i$), we have that $\tilde{\triangle}_i=0$ and $\tilde{\triangle}_i>0$ otherwise. Hence we will attempt to program $\sigma_i(t)$ to minimize $\tilde{\triangle}_i$. This is greatly facilitated if we choose $\nu$ so that
\begin{equation}
\tau_s < \nu^{-1} < \tau_N,
\end{equation}
 where $\tau_s$ is the time scale on which the node dynamics evolves (e.g., the time scale for the evolution of $x_i(t)$), and $\tau_N$ is the time scale on which the network evolves (i.e., the time scale on which $A_{ij}(t)$ and hence $\bar \sigma_i(t)$ change). With this assumption, $\sigma_i(t')$ in (\ref{tri}) can be replaced by $\sigma_i(t)$ to yield the following approximation to $\tilde{\triangle}_i$,
\begin{equation}
\triangle_i(t)= \sigma_i^2(t) B_i(t) - 2 \gamma \sigma_i(t) C_i(t) + \gamma^2 D_i(t), \label{tri2}
\end{equation}
where
\begin{equation}
B_i(t)=  \int^t e^{-\nu (t-t')} s^2_i(t') dt', \label{B}
\end{equation}
\begin{equation}
C_i(t)=  \int^t e^{-\nu (t-t')} s_i(t') \cdot H (x_i(t')) dt', \label{C}
\end{equation}
\begin{equation}
D_i(t)=  \int^t e^{-\nu (t-t')} (H (x_i(t')))^2 dt'. \label{D}
\end{equation}

Since $\tilde \triangle_i=0$ at synchronization and is positive otherwise, one option is to program $\sigma_i(t)$ so as to seek the minimum of $\triangle_i$ by the following gradient descent relaxation,
\begin{equation}
\frac{d \sigma_i(t)}{d t}= -\alpha \frac{d \triangle_i}{d \sigma_i}= -2 \alpha (\sigma_i B_i - \gamma C_i) \label{GD},
\end{equation}
where $\alpha$ is a parameter that determines the relaxation time scale and $\triangle_i(\sigma_i)$ may be viewed as a potential function for the gradient flow (\ref{GD}).
To eliminate the need for calculating the integrals (\ref{B}) and (\ref{C}) at each time step, we note that $B_i$ and $C_i$ satisfy the following first order differential equations,
\begin{equation}
\frac{d B_i(t)}{d t}= - \nu B_i +  s_i^2, \quad 
\frac{d C_i(t)}{d t}= - \nu C_i +  s_i \cdot H (x_i). \label{CC}
\end{equation}
%
Thus our adaptive strategy is described by the set of differential equations   (\ref{main},\ref{tri2}, \ref{CC}).

In order to test the above described strategy we have performed a series of numerical experiments that we now describe. In our initial experiments, we consider a random network of $N$ nodes and $<k>N/2$ links, where $<k>$ is the network average degree. At $t=0$ we assume that the adjacency matrix is $A_{ij}(0)=A_{ji}(0)=1$ if a link exists between $i$ and $j$, and $A_{ij}(0)=0$ otherwise. For $t>0$ we assume the following network evolution,
\begin{equation}
A_{ij}(t)=A_{ij}(0)(1+ \epsilon_{ij} \sin (\omega_{ij}t)),  \label{net_ev}
\end{equation}
where the $\epsilon_{ij}$ are random numbers drawn from a uniform distribution between $0$ and $1$ and the $\omega_{ij}$ are random numbers drawn from a uniform distribution between $\omega_{min}>0$ and $\omega_{max}>\omega_{min}$, where $\tau_N=(\omega_{max})^{-1}$ is much longer than the characteristic time scale of the dynamics at the network nodes $\tau_s$. 

As an example, we consider a network of coupled R\"ossler oscillators, $x_i=(x_{i1},x_{i2},x_{i3})^T$ and
$F(x_i)=[-x_{i2}-x_{i3}; x_{i1} + 0.165 x_{i2}; 0.2+ (x_{i1}-10 )  x_{i3}]$.
We choose the oscillators to be linearly coupled in the $x_{i1}$ variable, i.e., $H(x)=\mathbf{H} x$, where $\mathbf{H}=[ 1 \quad 0 \quad  0; \quad 0 \quad  0 \quad 0; \quad 0 \quad 0 \quad 0]$. We have also investigated other choices for $H(x)$ and obtained similar results.

 Here, for the sake of simplicity, we assume that the dynamics of the adaptation process (\ref{GD}) is fast. Thus taking $\alpha \rightarrow \infty$,  we have that $\sigma_i(t)$ rapidly converges to $\gamma C_i(t)/B_i(t)$, where the dynamics of $B_i(t)$ and $C_i(t)$ are given by (\ref{CC}), and in place of (\ref{tri2}) we use
\begin{equation}
\sigma_i= \gamma \frac{C_i(t)}{B_i(t)} \label{gcb}.
\end{equation}

We have found that the value of the parameter $\nu$ as well as the initial conditions on $C_i$ and $B_i$ can significantly impact the network behavior.  With respect to the initialization of $C_i$ and $B_i$ we emphasize that, while it may be physically difficult to initialize the variables $x_i$ in a prescribed way, in contrast, initializations of $B_i$ and $C_i$ can be freely specified, because $B_i$ and $C_i$ are internal variables that we use only in computing our adaptive changes. Assuming that $A_{ij}(0)$ is known, we set $C_i(0)= B_i(0) (\sum_j A_{ij}(0))^{-1}$ to satisfy (\ref{sigma}). We would like to choose $B_i(0)$ in such a way that, if our system consisting of Eqs. (\ref{main}, \ref{CC}, \ref{gcb}) has attractors other than the desired synchronism tracking solution $( \bar{A}_{ij}(t) \approx A_{ij}(t) )$, then these other spurious attractors do not capture the orbit. To promote this we wish to choose the $B_{ij}(0)$ so that the initial condition is likely to be in the basin of attraction of our desired solution. To this end, we assume that we are in a synchronized state and average the first of Eqs. (16) for $B_i(t)$ over the chaotic oscillations thus yielding $B_i \simeq <s_i^2>$. Noting the definition (\ref{si}) of $s_i$ and our choice $H(x)=(x,0,0)^T$ for our example we have $<s_i^2> \simeq <k^2> <x_{s1}^2>_t$, where $<k^2>$ is the second moment of the network degree distribution and $<x_{s1}^2>_t$ denotes a time average of $x_{s1}(t)$ for the synchronous chaotic dynamics, Eq. (\ref{xs}). Thus we choose
\begin{equation}
B_i(0)=<k^2> <x_{s1}^2>_{t}, \label{diciann}
\end{equation}
which yields for the example in Fig. 1, $B_i(0) \approx 10^4$.

\begin{figure}[t]
\centerline{\psfig{figure=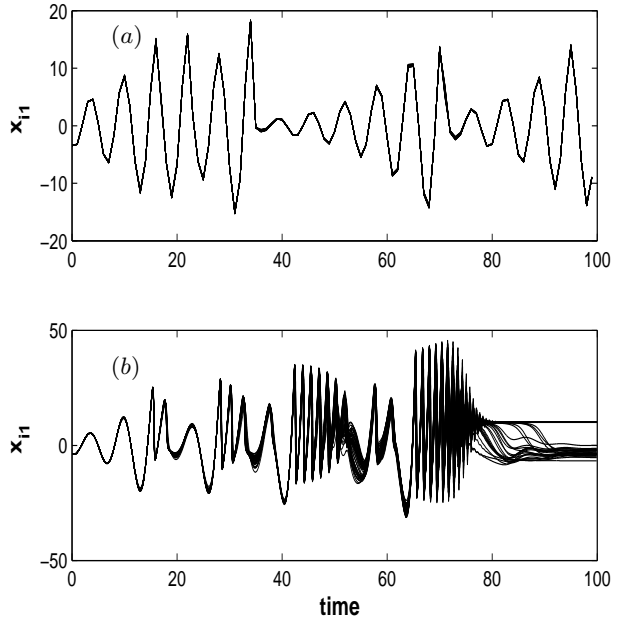,width=9cm,height=9cm}} 
\begin{picture}(0,0)(0,0)
\put(-80,235){\small ${(a)}$} \put(-80,110){\small
$(b)$}
\end{picture}
\caption{\small The plots show the time evolutions from $t=0$ to $t=100$ of $x_{i1}(t), i=1,..,N$,  while the network topology evolves according to (\ref{net_ev}) in the case where the adaptation (\ref{CC}, \ref{gcb}) was implemented (a) and in the case where it was not implemented (b). Here the oscillators start synchronized, $\nu=1/(2 \tau_s)$, $\gamma=2$.  The network parameters are as follows: $N=50$, $<k>=10$. \label{AD}}
\end{figure}

As a first experiment, we have sought to track the synchronous evolution for the evolving network, with 
the adaptive strategy described above.
We started from an initial condition in which all the oscillators are in the same state, $x_{i1}(0)=x_1^0, x_{i2}(0)=x_2^0, x_{i3}(0)=x_3^0$, $i=1,...,N$, where $x^0=(x_1^0,x_2^0,x_3^0)^T$ is a randomly chosen point belonging to the R\"ossler attractor and $\sigma_i(0)= \bar \sigma_i(0)$, $i=1,...N$, with $B_i(0)$ given by (\ref{diciann}) at each node $i$. We considered a network of $N=50$ nodes and average degree $<k>=10$. We took $\gamma=2$ so as to ensure the stability of the synchronous evolution in the case where the $A_{ij}$ are constant in time, $A_{ij}=A_{ij}(0)$. 
We assumed $\nu=1/(2 \tau_s)$, where we took $\tau_s$ to be the time at which the autocorrelation function of $x_s(t)$ obtained from numerical solution of (\ref{xs}) becomes $0.5$, $\tau_s \simeq 0.7$. The network topology was evolved as in (\ref{net_ev}), with $\omega_{min}=0.01$ and $\omega_{max}=0.02$.
Fig. \ref{AD}(a) shows superposed plots of the time evolutions of $x_{i1}(t), i=1,..,N$ from $t=0$ to $t=100$
for a case in which adaptation was implemented. We see from this figure that all the $N=50$ solutions evolve almost identically. Furthermore, their behavior is as described by the solution of the uncoupled chaotic dynamics, Eq. (\ref{xs}). 
In contrast, Fig. \ref{AD}(b) shows the same example but for the case in which adaption was not implemented (i.e., $\bar{\sigma}_i(t)=\sigma_i(0)$ for all time). In this case a synchronized solution obeying Eq. (\ref{xs}) is not attained, and after the network has significantly evolved ($t \geq 40$) there is appreciable spread amongst the $x_i$ at different network nodes. 

For the above experiment we found that the results were robust to changes in $B_i(0)$ from its nominal value of $10^4$ obtained from Eq. (19); e.g., using $B_i(0)=1$ gave essentially the same results. As we will see from our next numerical experiments, this is not always the case.
Fig. 1 shows that the proposed adaptive strategy technique is effective for tracking an initially synchronous state.
In what follows, we address the `global synchronization' problem for the network (\ref{main}), i.e., we consider initial conditions, which can be far from the synchronization manifold (\ref{si}). 
Indeed, in real situations it may often not be feasible to initialize with near identical states $x_i(0)$ on each node. In order to evaluate the effects of initial conditions $x_i(0)$ that differ from node to node,  we consider initializing the network as follows,
\begin{equation}
x_{i1}^0= x_1^0 + c \rho_1 \epsilon_{ix}, \quad  x_{i2}^0= x_2^0 + c \rho_2 \epsilon_{iy}, \quad  x_{i3}^0= x_3^0 + c \rho_3 |\epsilon_{iz}|, \label{IC}
\end{equation}
where $(x_1^0,x_2^0,x_3^0)$ is a randomly chosen point on the R\"ossler attractor;  $\epsilon_{ix}, \epsilon_{iy}$ and $\epsilon_{iz}$ are zero-mean independent random numbers of unit variance drawn from a normal distribution; $\rho_1=7.45,\rho_2=7.08,\rho_3=4.25$ are the standard deviations of the time evolutions of the states $x_s,y_s,z_s$ from numerical solution of (\ref{xs}) (calculated over a long time evolution); and $c$ is a parameter characterizing the degree to which the initial conditions vary from node to node.
\begin{figure}[t]
\centerline{\psfig{figure=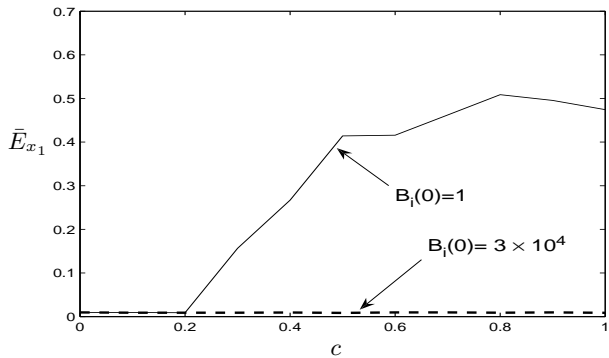,width=9cm,height=5cm}}
\begin{picture}(0,0)(0,0)
\put(-0,12){\small ${c}$} \put(-122,90){\small
${\bar{E}_{x_1}}$}
\end{picture}
\caption{\small $\bar{E}_{x_1}$ calculated between $t_1=500$ and $t_2=1000$, vs $c$;  $N=100$, $<k>=20$, $\nu=1/(2 \tau_s)$, $\gamma=2$, $B_i(0)=1$. The dashed line represents $\bar{E}_{x_1}$ vs $c$ for the case $B_i(0)=3 \times 10^4$.} \label{A3}
\end{figure}
We define an average synchronization error $\bar{E}_{x_1}$ for the evolution of the variables $x_{1i}(t)$ ($i=1,2,...,N$) 
\begin{equation}
\bar{E}_{x_1}=\frac{1}{N (t_2-t_1)\rho_1} {\int_{t_1}^{t_2} \sum_i{ |x_{i1}(t)-\bar{x}_{i1}(t)|} dt},
\end{equation}
where $\bar x_{i1}(t)=N^{-1} \sum_{i=1}^{N} x_{i1}(t)$ and $\rho_1=$ $<(x_{s1} - <x_{s1}>_t)^2>_t^{1/2}$, where $<...>_t$ indicates the time average and the subscript $s$ denotes evolution of $x=(x_1,x_2,x_3)^T$ in the synchronous state (i.e., using dynamics from Eq. \ref{xs}).
As shown by the dashed line in Fig. 2, obtained with $B_i(0) \simeq 3 \times 10^4$, given by (\ref{diciann}), we achieve good synchronization of the evolving adaptive network. In order to see the effect of an arbitrary less rational choice of $B_i(0)$ we have repeated this experiment using $B_i(0)=1$.
The solid line in Fig. \ref{A3}, shows $\bar{E}_{x_1}$, with $(t_1,t_2)=(500,1000)$, versus $c$, for $B_i(0)=1$. It is seen that as $c$ increases above about $0.2$, the network fails to synchronize.
In contrast, with $B_i(0)$ from (\ref{diciann}), synchronization was achieved for any value of the parameter $c$ between $0$ and $1$, thus illustrating the impact of properly choosing the initial $B_i$ value.
This is because, when the node states are not initialized closely enough, the network trajectories may be attracted by another attractor of the dynamical system (\ref{main},\ref{CC},\ref{gcb}), that is different from the synchronous attractor (\ref{si}). 

Finally, we have also tested the robustness of our scheme to deviations of the individual systems from identicality. To this end, we replace $F(x_i)$ in Eq. (\ref{main}) by $F_i(x_i)=[-x_{i2} -x_{i3}; x_{i1}+ 0.165 (1+ \Delta \delta_i); 0.2 +(x_{i1}-10) x_{i3}]$, where for each node $i$ the parameter $\delta_i$ is chosen randomly with uniform density in the interval $[-1,1]$ and repeat our original experiment (the experiment resulting in Fig. 1). 
The parameter $\Delta$ characterizes the degree of non-identicality of the node dynamical systems ($\Delta=0$ for Figs. 1 and 2).
Our results show, for example, that for $\Delta <0.2$, the synchronization error is less than $4 \%$, i.e., $\bar{E}_{x_1} \lesssim 0.04$, thus indicating that good results may still be obtained when the coupled systems deviate from being precisely identical.

In conclusion, we have shown that an adaptive strategy can be used for the tracking of synchronization of time evolving network systems whose network evolution is unknown at the nodes of the network. We have also evaluated the effects of variable initial conditions at the network nodes, and observed that global synchronization may be achieved if the initial conditions of the adaptive variables ($B_i(0),C_i(0)$) are chosen appropriately, i.e., from the basin of attraction of (\ref{main},\ref{CC},\ref{gcb}). Preliminary numerical experiments have shown the effectiveness of our proposed strategy in yielding approximate synchronization also for networks of non-identical  systems.

%
%




This work was supported by grants from ONR (N000140611004), NSF (PHY0456240) and by a DOD MURI grant (N000140710734).

\end{document}